\documentclass[aps,nofootinbib,superscriptaddress,floatfix,prd,twocolumn]{revtex4} 
\usepackage{graphicx,amsmath,amssymb,color,bm,xstring}
\usepackage{hyperref}

\def\be{\begin{equation}}
\def\ee{\end{equation}}
\def\bea{\begin{eqnarray}}
\def\eea{\end{eqnarray}}
\newcommand{\vs}{\nonumber\\}
\def\ba#1\ea{\begin{align}#1\end{align}}
\def\bg#1\eg{\begin{gather}#1\end{gather}}

\def\Mpl{M_\text{Pl}}

\newcommand{\refeq}[1]{Eq.~(\ref{eq:#1})}          
\newcommand{\refeqs}[2]{Eqs.~(\ref{eq:#1})--(\ref{eq:#2})}          
          
\newcommand{\reffig}[1]{Fig.~\ref{fig:#1}}          
          
\newcommand{\refsec}[1]{Sec.~\ref{sec:#1}}

\def\lapl{\nabla^2}

\renewcommand{\d}{\delta}

\newcommand{\Om}{\Omega_m}

\def\L{\mathcal{L}}

\newcommand{\comment}[1]{}

\def\Mpch{\,h^{-1}\,{\rm Mpc}}

\def\Lf{\Lambda^{-4}}

\begin{document}

\title{Monodromic Dark Energy}

\author{Fabian Schmidt}
\affiliation{Max-Planck-Institut f\"ur Astrophysik, Karl-Schwarzschild-Str.~1, 85748~Garching, Germany}

\begin{abstract}
  Since the discovery of the accelerated expansion of the Universe, the constraints on the equation of state $w_\text{DE}$ of dark energy, the stress-energy component responsible for the acceleration, have tightened significantly. These constraints generally assume an equation of state that is slowly varying in time. We argue that there is good theoretical motivation to consider ``monodromic'' scenarios with periodic modulations of the dark energy potential. We provide a simple parametrization of such models, and show that these leave room for significant, periodic departures of $w_\text{DE}$ from $-1$. Moreover, simple models with non-standard kinetic term result in interesting large-scale structure phenomenology beyond that of standard slow-roll dark energy. All these scenarios are best constrained in a dedicated search, as current analyses average over relatively wide redshift ranges.
\end{abstract}

\date{\today}

\maketitle

\section{Introduction}
\label{sec:intro}

Since the conclusive detection in 1998 \cite{Riess:1998cb,Perlmutter:1998np}, overwhelming evidence has accumulated that points towards
an accelerated expansion of the Universe which, in the context of
General Relativity (GR) implies the existence of a form of stress energy
with negative pressure, \emph{dark energy} \cite{Frieman:2008sn}. In particular, the dark energy equation of state $w_\text{DE} = p/\rho$ is now
constrained to be close to $-1$ \citep[e.g.,][]{JLA:2014,planck:2015,alam/etal:2016,DESy1} (see \cite{huterer/shafer:2017} for a recent review). A cosmological constant is the simplest
possibility, which is so far broadly consistent with all observational constraints coming from geometric and large-scale growth probes.
However, the value of the cosmological constant is, from a high-energy physics perspective, highly fine-tuned (see \cite{Weinberg:1988cp} for a review). An alternative is to make the cosmological constant dynamical, by promoting it to the potential energy $V(\phi)$ of a scalar field $\phi$, often dubbed \emph{quintessence} (see \cite{Copeland:2006wr} for a comprehensive, and \cite{joyce/lombriser/schmidt:2016} for a brief overview). If the potential is sufficiently flat, for example $V(\phi) \propto (\phi/\phi_0)^{-\alpha}$ with $\alpha \ll 1$, the field rolls slowly and is thus potential energy dominated, yielding a stress-energy contribution with equation of state close to $-1$ \cite{Wetterich:1987fm,Peebles:1987ek,ratra/peebles}. Moreover, during the matter-dominated epoch of the Universe, the field follows a tracking solution and thus reduces the need for fine-tuning of the model parameters.  

An approximately flat potential can be realized in a technically natural way by introducing a shift symmetry $\phi\to \phi+c$, which is then weakly broken. In the well-known explicit constructions of inflation in the context of string theory \cite{silverstein/westphal:2008,mcallister/etal:2010}, known as \emph{axion monodromy}, nonperturbative effects lead to periodic modulations of the potential. These are responsible for an array of interesting signatures \cite{chen/easther/lim,flauger/etal:2010,flauger/pajer}. Motivated by this fact, we study in this paper the phenomenological consequences of a dark energy potential of the form
\be
V(\phi) = C \left(\frac{\phi}{\phi_0}\right)^{-\alpha} \left [1 - A \sin (\nu\phi)\right]\,,
\label{eq:V}
\ee
where $A$ is the amplitude of the periodic modulation (with $|A|\ll 1$), while $\nu$ is the frequency in field space.\footnote{In axion monodromy, the smooth component of the potential is linear. We have generalized this to a power-law here, though the phenomonological consequences are not sensitive to the shape of the smooth potential.}  Apart from the shape of the smooth part of the potential (power-law vs. exponential), this is precisely the potential studied in \cite{dodelson/kaplinghat/stewart:2000}. It is different from the Pseudo-Nambu-Goldstone scenario of \cite{frieman/etal:1995} (see also \cite{kamionkowski/pradler/walker,damico/etal:2016}), where $V(\phi) = V_\star [1 + \cos(\phi/f)]$ is a non-monotonic potential, as \refeq{V} allows for the field to pass through multiple oscillations driven by the potential (several oscillations can however happen in the latter scenario if multiple decay constants $f_i$ are involved \cite{damico/etal:2016}).  
Further, this is a different physical setup than oscillations of a $U(1)$ quintessence field in a power-law potential \cite{spintessence1,spintessence2,spintessence3,spintessence4}, which have been shown to be unstable and hence not suitable for explaining a sustained period of acceleration \cite{johnson/kamionkowski}.

On the other hand, the scenario considered here has a wide range of parameter space which is stable (we discuss theoretical constraints in \refsec{thconst}). Moreover, the observable oscillatory features can have periods that are naturally much smaller than a Hubble time. Previous studies of oscillatory potentials \cite{barenboim1,barenboim2,linder:2006,kurek2}, and constraints using observational data \cite{xia/etal:2005,kurek1,kurek2,pace/etal:2011}, typically focused on periods of order a Hubble time. The window of rapidly oscillating dark energy model space thus remains largely unexplored. Since the assumption of a smooth evolution of the dark energy density is built into almost all observational constraints published so far, this leaves open the possibility of significant periodic deviations from an equation of state $w_\text{DE} \sim -1$.

Moreover, a canonical scalar field is not the only possible option. Models with non-standard kinetic term, referred to as \emph{k-essence}, can exhibit similar tracking behavior \cite{chiba/okabe/yamaguchi,ArmendarizPicon:2000ah}. Indeed, non-standard kinetic terms also appear in the context of string theory in form of the Dirac-Born-Infeld action \cite{silverstein/tong,alishahiha/etal:2004}. In these models, the sound speed is naturally small, so that perturbations in the dark energy density are not negligible. We will see that this leads to even more interesting signatures in large-scale structure (LSS): unlike the case of k-essence with a smoothly varying equation of state, which is observationally difficult to distinguish from quintessence, a periodic modulation of the form \refeq{V} produces signatures in the growth of structure which are accessible to galaxy redshift as well as weak lensing surveys. Again, this is a phenomenological window which currently is almost entirely unexplored. 

Throughout, we will assume a spatially flat background, and assume $\Omega_{m0} = 0.27 = 1 - \Omega_{{\rm DE},0}$ as fiducial values. We will also choose $\alpha=0.2$ as default, leading to a time-averaged equation of state of $\bar w_\text{DE} \approx -0.9$. The outline is as follows: \refsec{quintessence}--\ref{sec:kessence} present the dark energy models. \refsec{thconst} discusses theoretical constraints on the viable model space. We then derive the observable signatures of these models in \refsec{obs}, and conclude in \refsec{concl}.

\section{Monodromic quintessence}
\label{sec:quintessence}

Consider the following action for a canonical scalar field $\phi$ minimally coupled to gravity:
\be
S = \int d^4x \sqrt{-g} \left[ \frac12 \Mpl^2 R + \frac12 \nabla_\mu\phi\nabla^\mu\phi + V(\phi) + \L_m \right]\,,
\label{eq:actionQ}
\ee
where $V(\phi)$ is given by \refeq{V}, $\L_m$ is the matter action, and we assume no coupling between $\phi$ and other species apart from gravity. Again, this is essentially identical to the model proposed in \cite{dodelson/kaplinghat/stewart:2000}. The quintessence contribution to the stress-energy tensor is of the perfect-fluid form, $(T^Q)^\mu_{\ \nu} = \text{diag}(-\rho_Q, p_Q,p_Q,p_Q)$.  We consider a matter sector that consists of cold pressureless matter with equation of state $w_m=0$.  Restricting to a spatially homogeneous setting, in which the metric becomes of the Friedmann-Robertson-Walker (FRW) form, \refeq{actionQ} leads to the equation of motion 
\ba
\ddot\phi + 3 H \dot\phi + V_{,\phi} =\:& 0
\,,
\label{eq:eomQ}
\ea
where here and throughout, dots indicate derivatives with respect to time $t$. We will present numerical integration results of \refeq{eomQ} below. First, we begin with the case without oscillations.

Upon setting $A=0$ in \refeq{V}, we recover the well-known power-law potential first considered by \cite{ratra/peebles}. A power-law ansatz $\bar\phi(t) = \tilde\phi_0 t^p$, with $\tilde\phi_0 = \phi_0/t_0^p$, in matter domination where $H(t) = 2/(3t)$ yields
\be
p = \frac{2}{2+\alpha} \quad\mbox{and}\quad p^2 + p = \alpha C \tilde\phi_0^{-\alpha-2}\,.
\label{eq:Qtracking}
\ee
The equation of state and energy density of this tracking solution are
\ba
w_Q =\:& \frac{p_Q}{\rho_Q} = \frac{\dot\phi^2/2 - V(\phi)}{\dot\phi^2/2 + V(\phi)} \vs
\stackrel{\text{tracking}}{=}\:& \frac{-1 + \alpha^4/4 + \alpha^3/16}{1 + \alpha/2+\alpha^2/4 + \alpha^3/16}
\,.
\ea
\begin{figure}[t!]
  \centering
\includegraphics[width=0.49\textwidth]{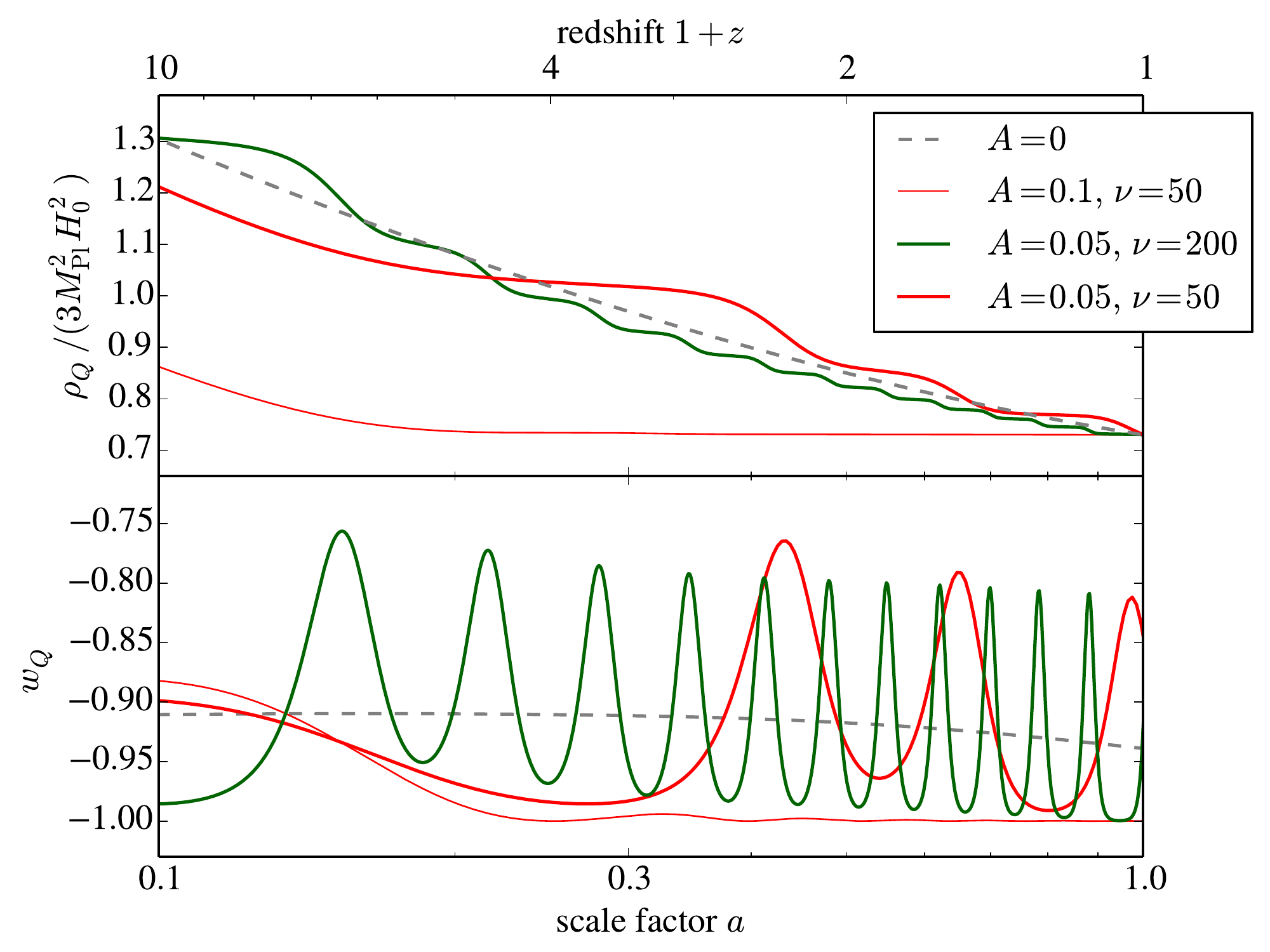}
\caption{Evolution of the energy density ($\rho_Q$, top) and equation of state ($w_Q$, bottom) of the monodromic quintessence model. The frequency $\nu$ is given in units of $\Mpl^{-1}$.
\label{fig:rhoQ}}
\end{figure}
We see that $w_Q$ is constant during matter domination, and that $w_Q$ approaches $-1$ as $\alpha\to 0$. In the following, we will choose $\alpha=0.2$ as fiducial value, leading to $w_Q=-0.89$ on the tracking solution. In the presence of oscillations, \refeq{eomQ} is no longer solvable in closed form. However, if we perturb around the tracking solution $\bar\phi$ by writing
\be
\phi(t) = \bar\phi(t) + \varphi(t)\,,
\ee
and work to linear order in the amplitude of oscillations $A$, we obtain in the limit $\alpha\to 0$ during matter domination: 
\ba
\varphi(t) =\:& \frac{\text{const}}{t} \exp\left[\pm i \omega t \right]\,,\quad\mbox{with} \vs
\omega =\:& \sqrt{C A \nu^2 \sin[\nu\bar\phi]}\,.
\label{eq:varphiQ}
\ea
Thus, if the tracking solution is initialized with a small positive $\bar\phi$, $\varphi$ oscillates around $\bar\phi$ with a frequency $\omega$.  If $\nu \gg 1/\Mpl$, the field can oscillate many times within a Hubble time.

We next turn to the full numerical solution of the system. First, the evolution of the quintessence energy density can be obtained from the continuity equation,
\be
\dot\rho_Q + 3H (1+w_Q) \rho_Q = 0\,,
\label{eq:contQ}
\ee
while the Hubble rate is determined from the Friedmann equation:
\be
H^2 = \frac{1}{3\Mpl^2} \left[\rho_m + \rho_Q \right]\,,
\label{eq:Fr1}
\ee
where $\rho_m(t) \propto a^{-3}(t)$ and we have assumed a flat background. 
\reffig{rhoQ} shows the evolution of the quintessence equation of state and energy density both with and without oscillations. For this, we have integrated \refeq{eomQ} together with \refeq{Fr1}, using initial conditions at a scale factor $a \approx 10^{-4}$ given by the tracking solution \refeq{Qtracking}. We adjust the initial value of $(\phi/\phi_0)$ [where $C$ is determined from \refeq{Qtracking}] in order to obtain the desired value of $\Omega_{{\rm DE,}0} = \rho_Q(t_0)/(3\Mpl^2 H_0^2) = 0.73$.

We show results for two different choices of the oscillation frequency $\nu$ in field space, both with amplitude $A=0.05$. 
Note that the oscillations in the energy density are substantially smaller than those in $w_Q$, as $\rho_Q$ is given by a time integral over $w_Q$ [\refeq{contQ}]. We also show an example with a slightly larger amplitude $A=0.1$. In this case, the quintessence field becomes trapped in a local minimum of the potential, leading to $w_Q=-1$ (see, e.g. \cite{deputter/linder:2008}). Indeed, we see that if
\be
| A \phi_0 \nu | \gtrsim 1\,,
\ee
the potential \refeq{V} is no longer monotonic. Since quintessence cannot cross the phantom divide to $w_Q < -1$, the evolution of the field stops once it has reached this value. Since, on the other hand, the average dark energy equation of state is observationally constrained to $\bar w_\text{DE} \lesssim -0.9$, this leaves a limited parameter space of viable monodromic quintessence models, with $|A| \lesssim 0.05$.

Note that, while the mass of the field $\phi$ is in general larger than $H$ in the monodromic quintessence scenario, the perturbations to the energy density of quintessence remain negligible. This is because the sound speed, defined in \refeq{wcs} below, is identical to unity in this model, so that pressure perturbations prevent subhorizon quintessence perturbations from growing. This is qualitatively different in the alternative scenario which we turn to next, which also has a significantly enlarged observationally allowed parameter space.

\section{Monodromic k-essence}
\label{sec:kessence}

We now generalize the action \refeq{actionQ} to a non-standard kinetic term, commonly referred to as k-essence, and consider the following action:
\be
S = \int d^4x \sqrt{-g} \left[ \frac12 \Mpl^2 R + p(\phi,X) + \L_m \right]\,,
\label{eq:actionK}
\ee
where the kinetic term is defined through
\be
X \equiv - \frac12 \Lf \nabla_\mu\phi\nabla^\mu\phi\,,
\label{eq:Xdef}
\ee
such that $X>0$. 
We have introduced the scale $\Lf$ to make $X$ dimensionless. At the background level, $X = \Lf\dot\phi^2/2$. While this scale is arbitrary, since it can be absorbed by a field redefinition, we choose the most natural value of $\Lambda = \sqrt{H_0 \Mpl}$. 
The stress-energy tensor for this field is still of the perfect-fluid form, with pressure and energy density given by
\ba
p_K =\:& p(\phi,X) \vs
\rho_K =\:& 2 X p_{,X}(\phi,X) - p(\phi,X)\,.
\label{eq:rhoK}
\ea
This immediately yields the equation of state $w_K$. Interestingly, the non-standard kinetic term yields a sound speed of the field, defined as the derivative of $p_K$ with respect to $\rho_K$ at fixed field value, which in general is different from unity:
\ba
c_s^2 \equiv \frac{\partial p_K}{\partial\rho_K}\Big|_{\phi}
= \frac{p_{K,X}}{\rho_{K,X}} = \frac{p_{,X}}{p_{,X} + 2 X p_{,XX}}\,.
\label{eq:wcs}
\ea
To avoid tachyonic behavior, we require $p_{,XX}>0$.  The equation of motion of $\phi$ at the background level can be compactly phrased in terms of $X = \Lf\dot\phi^2/2$, yielding
\be
(p_{,X} + 2 X p_{,XX}) \dot X + \sqrt{2X} (2 X p_{,X\phi} - p_{,\phi}) + 6 H X p_{,X} = 0\,.
\label{eq:eomX}
\ee
Following \cite{chiba/okabe/yamaguchi}, we choose
\be
p(\phi,X) = V(\phi) \left[- X +  X^2 \right]\,,
\label{eq:p-k}
\ee
where $V(\phi)$ is given by \refeq{V}. Note that $V(\phi)$ is not a potential here, but determines the amplitude of the kinetic term.  The negative pressure is obtained through the non-canonical kinetic term. 
Any function of the form $K(\phi) X + L(\phi) X^2$ (with $L \neq 0$) can be brought into this form through a field redefinition \cite{chiba/okabe/yamaguchi}.
  
Inserting \refeq{p-k} into \refeqs{rhoK}{wcs}, we obtain
\be
w(X) = \frac{1-X}{1-3X}
\quad\mbox{and}\quad
c_s^2(X) = \frac{1-2X}{1-6X}\,.
\label{eq:wcs2}
\ee
A cosmological constant behavior is attained in the limit $X\to 1/2$.  In close analogy to the power-law quintessence case, the k-essence model in the absence of oscillations ($A=0$) admits a scaling solution in matter domination, where $X = \bar X$ is constant and $\phi = \sqrt{2\bar X}\Lambda^2 t + \text{const}$. This scaling solution has
\be
\bar X(\alpha) = \frac{4-\alpha}{8-3\alpha}\,.
\label{eq:Xscaling}
\ee
As in the quintessence case, setting $\alpha=0$ yields a cosmological constant with $\bar X=1/2$ and $w_K=-1$; our fiducial choice in the following, $\alpha=0.2$, yields $w_K=-0.9$.

\begin{figure}[t!]
  \centering
\includegraphics[width=0.49\textwidth]{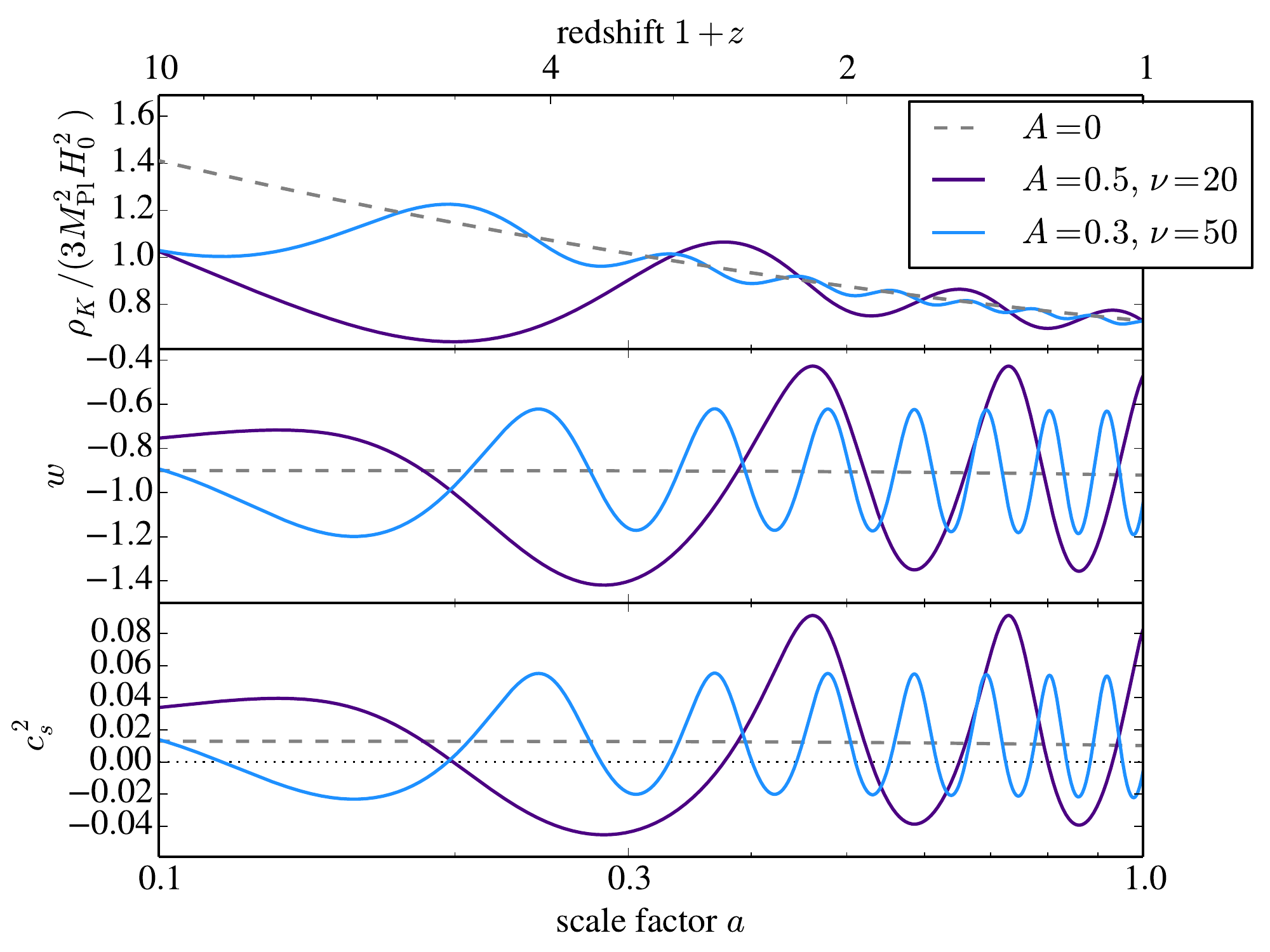}
\caption{Evolution of the energy density ($\rho_K$, top), equation of state ($w_K$, middle), and sound speed squared (bottom) of the monodromic k-essence model.
 The frequency $\nu$ is given in units of $\Mpl^{-1}$. \label{fig:rhoK}}
\end{figure}

Before turning to the results from the numerical integration of the full equation of motion \refeq{eomX}, let us again derive a simpler equation at linear order in the amplitude $A$, by expanding around the scaling background.  Writing
\be
\phi = \sqrt{2\bar X}\Lambda^2 t + \varphi(t)\,,
\ee
and taking the limit $\alpha\to 0$, we obtain
\be
\ddot\varphi(t) + \frac2t \dot\varphi(t) + \frac18 A \,\Lambda^2 \nu \cos\left(\Lambda^2 \nu t \right) = 0\,.
\ee
This equation can be straightforwardly solved to yield, in the limit of large $\nu t$,
\be
\varphi(t) =  - A \frac{\Lambda^2}{8\nu} \cos\left(\Lambda^2\nu t\right)\,.
\label{eq:varphiK}
\ee
Correspondingly, all of $\rho_K$, $X$, $w_K$ and $c_s^2$ oscillate around the values
given by the scaling solution, with approximately
\be
\frac{\Lambda^2 \nu}{2\pi H} \approx \frac{\nu\Mpl}{2\pi}
\label{eq:Nosc}
\ee
oscillations per Hubble time. While the amplitude of oscillations in $\rho_K$, and hence the Hubble rate $H$, are controlled by the field amplitude and are hence of order $A/\nu$, the oscillations in $w_K$ and $c_s^2$ are proportional to $\dot\varphi$, which oscillates with amplitude $A$. This will become relevant in the observable effects considered in \refsec{obs}. As shown in \cite{creminelli/etal:2009}, no pathologies arise in k-essence when crossing the phantom divide $w_K=-1$, as long as the sound speed vanishes at the crossing, and as long as gradient instabilities are countered by higher-derivative terms neglected in the action \refeq{actionK}. The former is precisely what happens in monodromic k-essence, as is clear from \refeq{wcs2} (see also \reffig{rhoK}). We will discuss in \refsec{thconst} whether brief episodes of tachyonic behavior can be allowed. 
Note that, by appropriate choice of $\alpha$ and $A$, one can ensure that $c_s^2 > 0$ always. This however restricts the allowed range of oscillation amplitudes significantly.

\reffig{rhoK} shows the evolution of the k-essence energy density and equation of state as well as sound speed squared obtained from a numerical integration of \refeq{eomX} [which is quite close to the result of the analytical approximation \refeq{varphiK}]. The initial conditions are again taken from the tracking solution at $a \approx 10^{-4}$, and the constant $C$ in $V(\phi)$ is adjusted to obtain the desired value of $\Omega_{{\rm DE,}0}=0.73$. Qualitatively, the behavior is similar to the monodromic quintessence case. However, there are several important differences. First, the oscillations are not damped [compare \refeq{varphiK} with \refeq{varphiQ}]. Second, the k-essence field can cross the phantom divide, allowing for significantly larger oscillations even for an average equation of state that is close to $-1$. Recall that $V(\phi)$ is not a potential in this model, and so the field does not get stuck even if $V$ is non-monotonic. Third, the k-essence model exhibits an oscillatory sound speed with $|c_s^2| \ll 1$, while $c_s^2=1$ always holds for quintessence.

\section{Theoretical constraints}
\label{sec:thconst}

We now briefly review theoretical constraints on the parameter space of the monodromic dark energy models considered here. We focus on constraints which are independent of the microscopic physics that leads to the potential \refeq{V}, and discuss the latter at the end of this section. In the quintessence case, we have already found the requirement $| A \phi_0 \nu | < 1$ in order to ensure a rolling field; otherwise, the field gets trapped in a local minimum, leading to an effective cosmological constant, which is uninteresting phenomenologically. No such constraint exists for the k-essence case.

However, unlike the monodromic quintessence case where $c_s=1$, in the k-essence case we have to confront the issue of gradient instabilities \cite{vikman:2004}. Within the comoving sound horizon of the k-essence field, defined as
\be
R_s(a) \equiv \frac{|c_s(a)|}{a H} \approx 150 \Mpch \left(\frac{|c_s(a)|}{0.05} \right)\,,
\label{eq:Rs}
\ee
an additional effective pressure force becomes relevant in the dynamics of the k-essence fluid. In particular, it sources a relative velocity divergence $\theta_{mK} \equiv \partial_{x,i} (v_K^i - v_m^i)$ between k-essence and matter, where $\partial_x$ denotes a derivative with respect to comoving coordinates, whose evolution equation is given by
\ba
\dot\theta_{mK} + H \theta_{mK} + \frac{c_s^2}{1+w_K} a^{-1} \lapl_x \d_K =\:& 0\,,
\label{eq:thetamk}
\ea
where $\d_K$ is the fractional energy density perturbation in the k-essence component and we have assumed $|c_s^2| \ll 1$. 
If $c_s^2 < 0$, this leads to an exponential growth instability in the dark energy component, known as gradient instability, as $\d_K$ is itself sourced by $- \theta_{mK}$. We can formally integrate \refeq{thetamk} to yield a stability constraint given by 
\be
\int_0^{\ln a} d\ln a' H^{-1}(a') D_K(a') \frac{c_s^2(a')}{1+w_K(a')} > 0\,,
\label{eq:stabconst}
\ee
where $D_K(a) = \d_K(k,a)/\d_K(k,1)$ is the k-essence growth factor which is in general scale dependent and a complicated function of time. Let us first consider large-scale perturbations whose time scale $1/\omega = 1/k$ is longer than one oscillation period of the field. From \refeq{Nosc}, this implies
\be
\frac{k}{a H} \lesssim \frac{\nu \Mpl}{2\pi}\,.
\ee
For these perturbations, the episodes of tachyonic behavior are too short to allow instabilities to grow. We have verified this by evaluating \refeq{stabconst} for the k-essence model introduced in the previous section, using the result for the k-essence perturbation obtained from \refeq{EPK} below, and find that the stability constraint is satisfied for a wide range of parameter space, including the regime where $A$ is of order 1. Fundamentally, this is because $c_s$ oscillates around a positive value in these models, and the combination $c_s^2/(1+w_K)$ is in fact always positive.

This argument no longer applies to small-scale perturbations with $k/aH \gg \nu\Mpl/2\pi$. These perturbations can in principle grow to become nonlinear within a single epoch of $c_s^2 < 0$. As argued in \cite{creminelli/etal:2009}, this instability can be countered by adding higher-derivative terms to the action \refeq{actionK}, for example $\bar\Lambda^{-2} (\Box\phi)^2$. These higher-derivative terms also control the cutoff of the effective description; that is, the model no longer provides controlled predictions in the effective field theory sense on spatial scales that are smaller than $\bar\Lambda^{-1}$. Since the episodes of tachyonic behavior are only brief, of order $(\nu\Mpl)^{-1}$ Hubble times, the requirement on the scale $\bar\Lambda$ associated with the higher-derivative term relaxes by a factor of $(\nu\Mpl)^{-1}$. Nevertheless, the k-essence model considered here either requires a very low cutoff (compared to local measurements of gravity), or significant fine-tuning between the natural scale $\Lambda = \Mpl^2 H_0^2$ and the much higher scale $\bar\Lambda$ of the higher-derivative terms.

The nontrivial sound speed in the k-essence model is induced by the non-canonical kinetic term, $\propto X + X^2$. This raises the question of whether higher-order terms in this expansion should be included, especially as $X\approx 1/2$ is not small. However, it turns out that \emph{any} k-essence model which admits a tracking solution with $X = $~const can be approximated by the form \refeq{p-k} as long as $X$ does not deviate strongly from its value on the tracking position. To see this, one can simply insert a general Lagrangian of the form $p(\phi,X) = V(\phi) K(X)$ into \refeq{eomX} and impose the tracking ansatz. Expanding $K(X)$ around the tracking point $X_{\rm tr}$, one finds that higher-order corrections to \refeq{p-k} scale as powers of $(X-X_{\rm tr})$. Now, even in the model with the strongest oscillations considered here, $A=0.5$, we find that $|X-X_{\rm tr}| < 0.1$ always. Thus, as long as the smooth component of $|1+w|$ is in the range allowed by observations, \refeq{p-k} is expected to represent the entire class of tracking k-essence models with an oscillatory potential. Of course, these statements are specific to the class of k-essence models which feature a tracking solution. 

Finally, we turn to somewhat more model-dependent constraints. First, we have neglected all higher-derivative operators in the dark energy Lagrangians. As discussed above, these higher-derivative operators are enhanced by $\omega/H$ in the monodromic case, compared to slowly-rolling scenarios, where $\omega$ is the oscillation frequency of the field in the cosmological solution, given by \refeq{varphiQ} in case of quintessence and $\omega/H \approx \nu \Mpl$ in the k-essence case. For the models considered here, this enhancement is as large as an order of magnitude. Whether such an enhancement makes higher-derivative operators relevant, depends on the microscopic physics responsible for the monodromic potential. In the context of inflation from axion monodromy realized in string theory, which of course happens at an energy scale much higher than dark energy, Ref.~\cite{flauger/etal:2010} found that higher-derivative terms remain suppressed as long as $\nu \Mpl$ is smaller than several thousand, and the oscillation amplitude is of order unity or less.

\section{Observables}
\label{sec:obs}

\begin{figure}[t!]
  \centering
\includegraphics[width=0.49\textwidth]{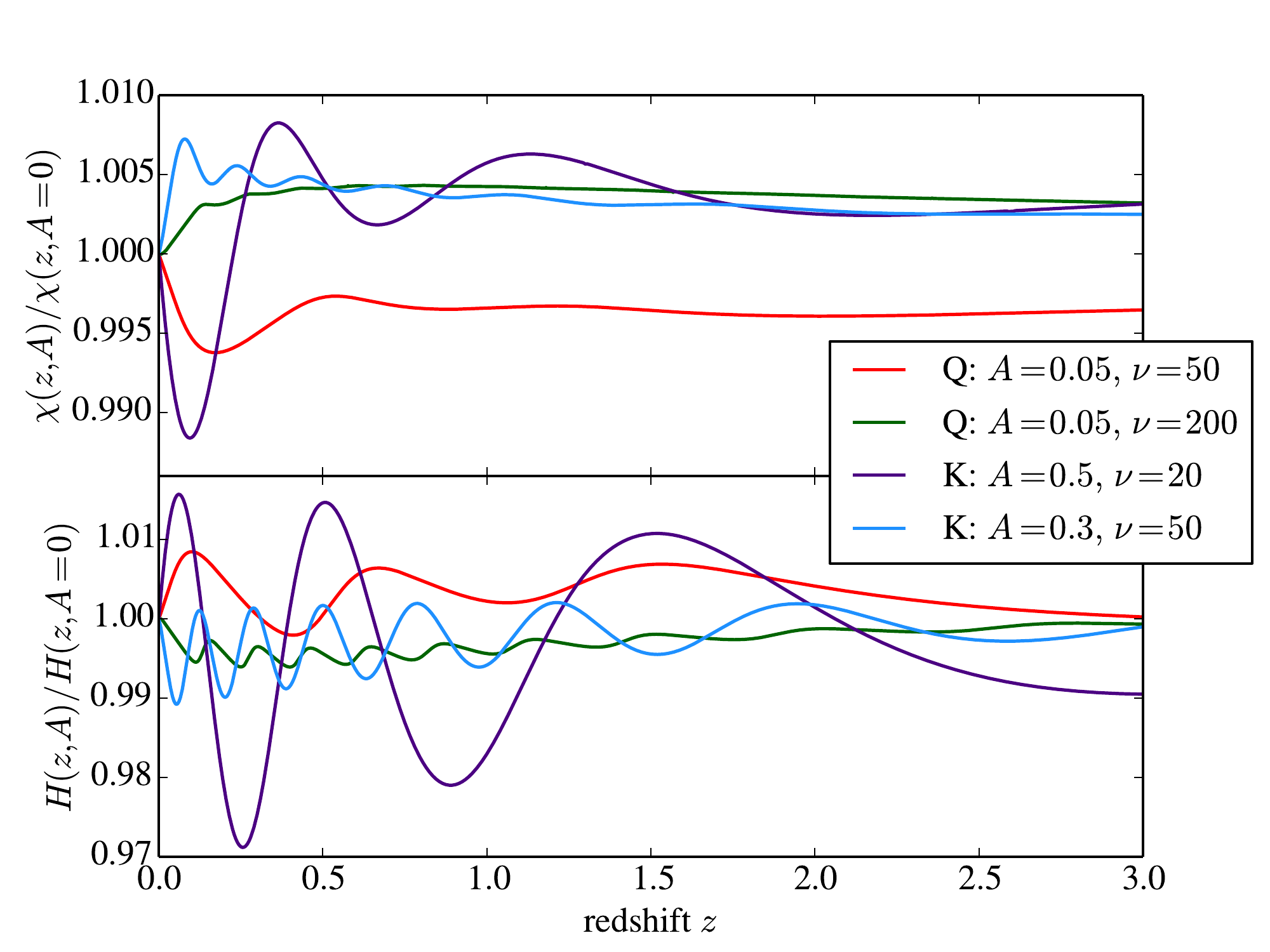}
\caption{Background observables in monodromic dark energy models: comoving distance $\chi(z)$ and Hubble expansion rate $H(z)$. In all cases, we show the ratio of oscillating models ($A > 0$) to the corresponding model without oscillations ($A=0$). 
\label{fig:obsbg}}
\end{figure}

We now consider the observable signatures of the monodromic quintessence and k-essence models. The simplest observables are distances, which are probed by standard candles such as type Ia supernovae, and standard rulers such as the baryon acoustic oscillation (BAO) feature in galaxy clustering. Given the assumed flat geometry, cosmological distances satify the simple relations
\ba
\chi(z) =\:& (1+z) d_A(z) = \frac{d_L(z)}{1+z}
= \int_0^z \frac{dz'}{H(z')}\,,
\ea
where $\chi$ is the comoving distance, $d_A$ is the angular diameter distance while $d_L$ is the luminosity distance. The distances correspond to one integral over $H \propto \rho_{\rm DE}$; thus, the signature of oscillations is even weaker in the distances than in the density, as can be seen in \reffig{obsbg} (see also \cite{linder:2006}).  Here and throughout, we show the ratio of predicted observables in a monodromic model with $A>0$ to the corresponding model with $A=0$. This is because we are interested in the oscillatory features as a signal, while smoothly varying changes in the observables could be explained by any standard dark energy model, for which the standard parametrizations through, for example, $w_0,w_a$ [\refeq{wparam}], are sufficient. 

Crucially, by sufficiently fine binning in redshift, standard candles can also probe the derivative ${\rm d}\chi/{\rm d}z = 1/H(z)$. Similarly, the BAO feature along the line-of-sight direction, as well as Alcock-Paczy\'nski (AP) distortions, probe $1/H(z)$. As shown in \reffig{obsbg}, the oscillations in $H(z)$ are significantly stronger than those in the distance. Moreover, the monodromic k-essence model shows much stronger oscillatory features than the quintessence case, a consequence of the limit on the amplitude of modulations in the quintessence case discussed at the end of \refsec{quintessence}. The signatures of monodromic k-essence are already accessible to current probes, as we will discuss in \refsec{concl}. Note however that when the radial BAO or AP scale are averaged over a wide redshift range, the oscillation signal is strongly diluted again. Thus, \emph{a dedicated analysis is necessary to obtain optimal constraints on monodromic dark energy models.} 

\begin{figure}[t!]
  \centering
\includegraphics[width=0.49\textwidth]{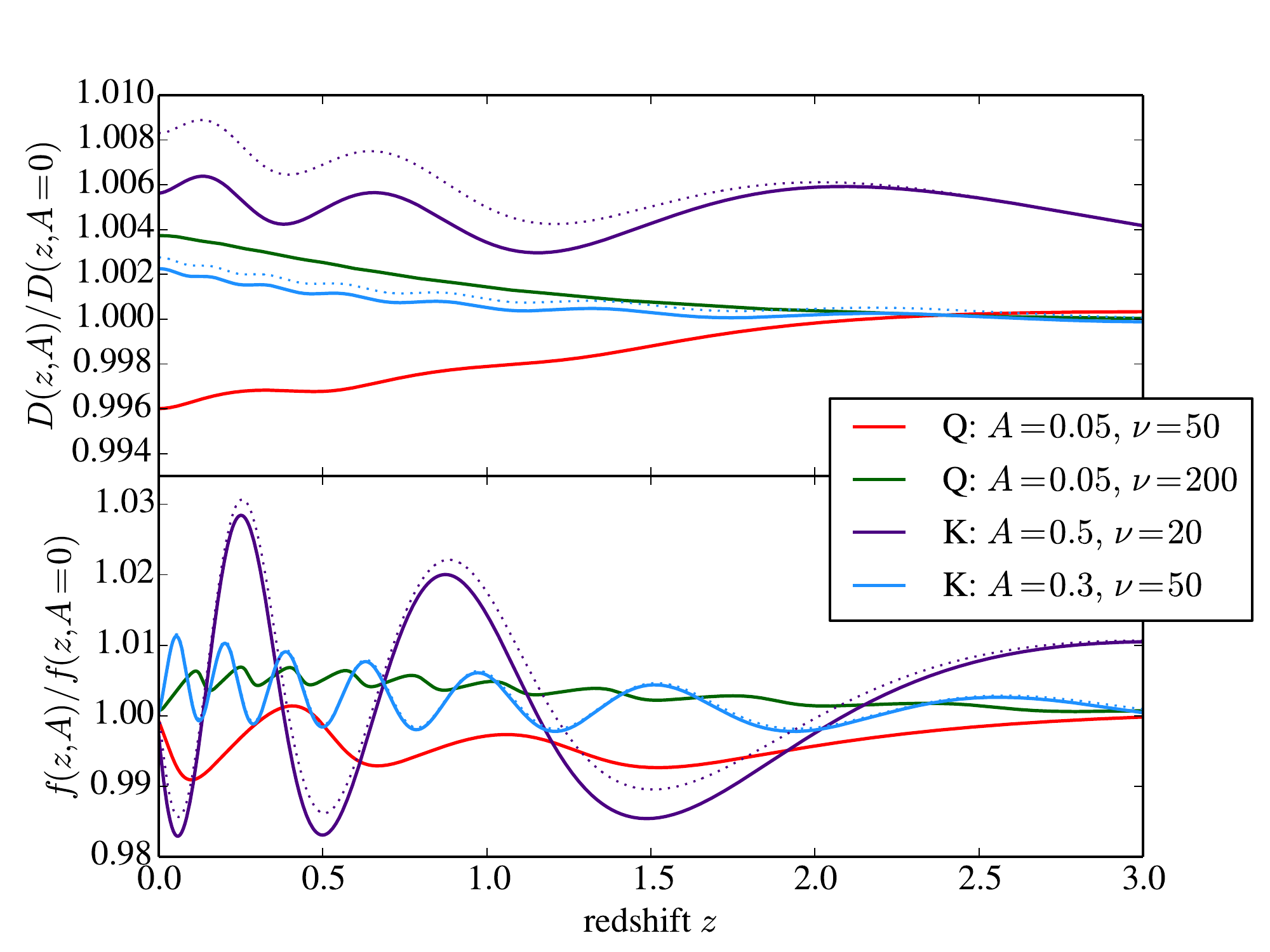}
\caption{Linear growth-of-structure observables in monodromic dark energy models: linear matter growth factor $D(z)$, and growth rate $f(z) d\ln D/d\ln a$. In all cases, we show the ratio of oscillating models ($A > 0$) to the corresponding model without oscillations ($A=0$). The dotted lines show the predictions in the k-essence case when neglecting dark energy perturbations. 
\label{fig:obsgrowth}}
\end{figure}

The same holds for the monodromy signatures in the growth of structure which we consider now, beginning with the
quintessence case. Since the sound horizon of the field is equal to the
comoving horizon, fractional perturbations in the dark energy density are less than $10^{-4}$ and can be neglected, as is usually done in studies of the growth of structure in dark energy cosmologies. The growth of matter perturbations at linear order is governed by the linearized Euler-Poisson system:
\ba
\dot\d_m + a^{-1} \theta =\:& 0 \vs
\dot\theta + H \theta - \frac32 a H^2  \Om(t) \d_m  =\:& 0\,.
\label{eq:EP}
\ea
These can be combined into a single equation for the linear growth factor
\be
\ddot D + 2 H \dot D - \frac32 \Omega_{m0} H_0^2 a^{-3} D = 0\,.
\ee
This relation holds in all models of dark energy where perturbations in the dark energy density can be neglected. 
We see that the growth factor corresponds to two integrals over the energy density and Hubble rate, and thus expect small imprints of oscillations in the growth factor. However, the large-scale clustering of galaxies also receives contributions from the velocity, induced by redshift-space distortions (RSD) \cite{kaiser:1987}. From \refeq{EP} we have $\theta = - a \dot\d_m = -a H f \d_m$, where $f=d\ln D/d\ln a$ is the linear growth rate. As seen in \reffig{obsgrowth}, the signatures in the growth rate are comparable to those in the Hubble rate. 

The monodromic k-essence model has an even more interesting phenomenology for LSS. Due to the nontrivial sound speed in this model, the comoving sound horizon $R_s$ [\refeq{Rs}] is much smaller than $(aH)^{-1}$.  
Within the sound horizon, perturbations in the dark energy are suppressed, although the oscillations in the sound speed might lead to interesting behavior even in this regime (see \refsec{thconst}). In the following, we consider perturbations which are much larger than $R_s$, where the field perturbations are unsuppressed. Equivalently, our predictions assume the limit $c_s^2 \to 0$, which Ref.~\cite{creminelli/etal:2009} argue should be used for any healthy model which cross the phantom divide $w_K=-1$. Then, the dark energy comoves with matter, and the linearized Euler-Poisson system becomes \cite{bean/dore,sefusatti/vernizzi}
\ba
\dot\d_m + a^{-1} \theta =\:& 0 \vs
\dot\d_K - 3 H w_K \d_K + a^{-1} (1+w_K) \theta =\:& 0 \vs
\dot\theta + H \theta - \frac32 H^2 a \left[ \Om(t) \d_m + \frac{\rho_K(t)}{3\Mpl H^2} \d_K\right] =\:& 0\,,
\label{eq:EPK}
\ea
where $\theta$ is the velocity divergence, while $\d_K$ is the fractional perturbation in the dark energy density, as in \refsec{thconst}. 

\begin{figure}[t!]
  \centering
\includegraphics[width=0.49\textwidth]{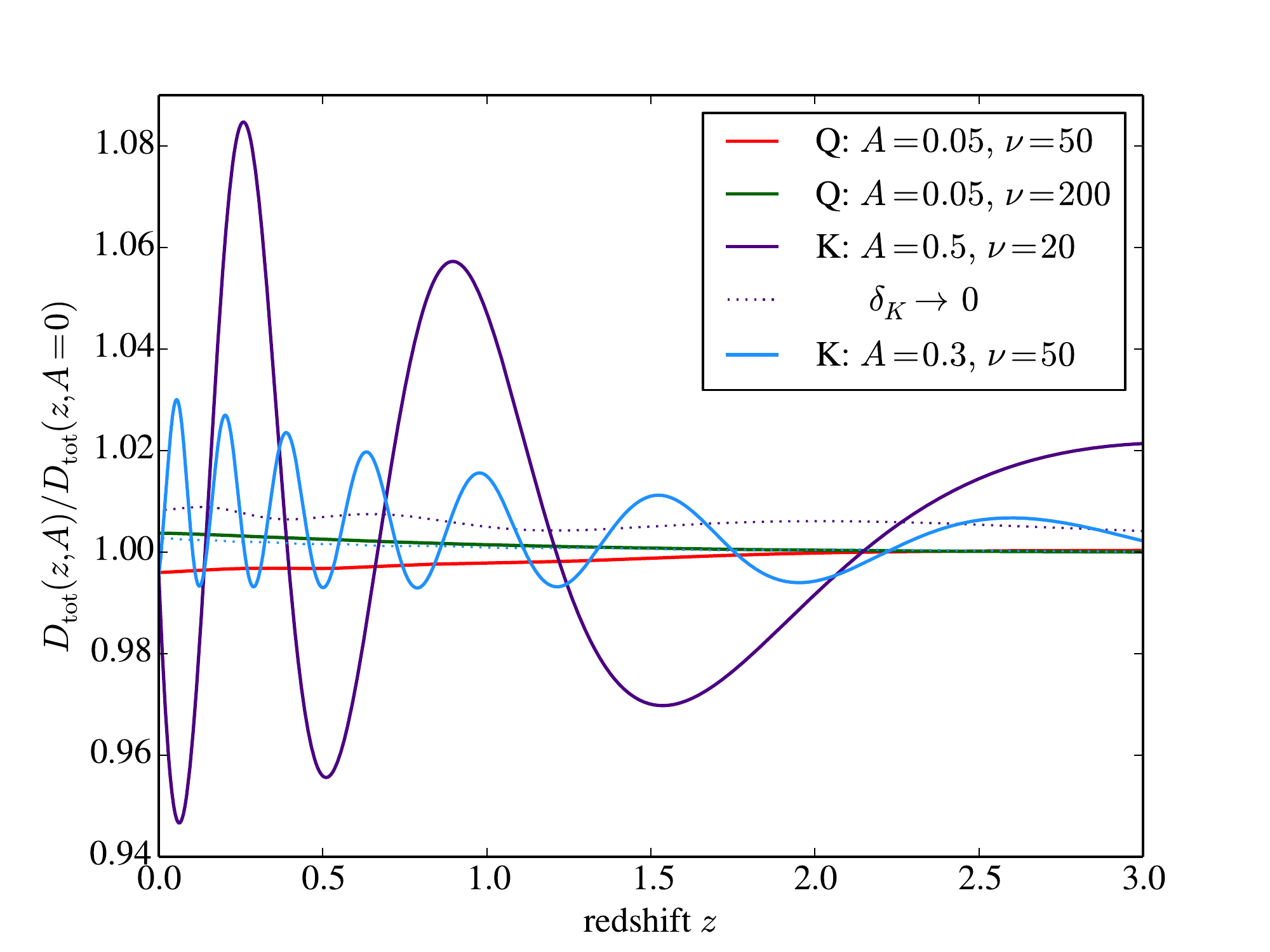}
\caption{Growth factor of the total stress-energy perturbation $\d_{\rm tot}$ defined in \refeq{dtot}, which determines gravitational lensing. For quintessence, this is the same as the matter growth displayed in \reffig{obsgrowth}. Shown is the ratio of oscillating models ($A > 0$) to the corresponding model without oscillations ($A=0$). The dotted lines show the predictions in the k-essence case when neglecting dark energy perturbations. 
\label{fig:obsgrowthtot}}
\end{figure}

The resulting growth factor and growth rate are shown in \reffig{obsgrowth}.\footnote{We integrate \refeq{EPK} assuming adiabatic initial conditions for matter and k-essence, which corresponds to the fastest growing mode.} As in the case of the background observables, the signal of oscillations is much stronger in the k-essence case due to the larger amplitudes of the potential modulation that is possible.  They can easily reach 5--10\% and are thus already accessible to existing data sets, as we will discuss in \refsec{concl}; however, again a dedicated analysis should be performed to search for these particular rapidly varying signatures.  Moreover, the dotted lines in \reffig{obsgrowth} show the result obtained when neglecting the dark energy density perturbations in the growth factor and growth rate, equivalent to integrating \refeq{EP} instead of \refeq{EPK}. Clearly, the effects are not negligible. The monodromic k-essence case thus offers an avenue to detect dark energy perturbations, which are challenging to detect for smooth equations of state (e.g., \cite{takada:2006}). 

So far, we have considered the growth of the matter sector which (on the large cales considered here) includes cold dark matter and baryons. However, large-scale strucutre also offers the opportunity to measure the total density perturbation
\be
\d_{\rm tot} \equiv \Om(t) \d_m + \frac{\rho_K(t)}{3\Mpl H^2} \d_K = \left(\frac32 a^2 H^2\right)^{-1} \lapl_x\Phi\,,
\label{eq:dtot}
\ee
which sources the gravitational potential $\Phi$. Here, we again consider scales larger than $R_s$, so that pressure perturbations in the dark energy can be neglected. In both models considered, as in almost all dark energy models, there is no anisotropic stress. This means that the two spacetime potentials are equal, and $\d_{\rm tot}$ can be probed directly through gravitational lensing (see, e.g. \cite{joyce/lombriser/schmidt:2016}). The prediction for the growth factor of $\d_{\rm tot}$ is shown in \reffig{obsgrowthtot}. In the quintessence case, it is identical to the matter growth. However, for k-essence, we see significantly stronger features than in the matter growth factor, which go up to a $\gtrsim 10\%$ change in the growth factor in one of the models considered. This strong signature exists because the dark energy density perturbation now contributes directly to the observable, rather than only through its gravitational backreaction on the matter growth; moreover, the oscillations in the growth of the dark energy perturbations are much stronger due to the effect of the oscillating equation of state $w_K$ [second line in \refeq{EPK}]. Note that, while they show order unity oscillations, the fractional dark energy perturbations remain small at all times (less than 10\% of the fractional matter density perturbations). This holds in the monodromic k-essence scenario as long as $A(1+\bar w_K) \ll 1$. 

Such a correlated, but different modulation in the growth of matter and gravitational lensing is a telltale signature of dark energy perturbations.

\section{Conclusions}
\label{sec:concl}

The main aim of this paper is to point out that there exists a significant theoretically motivated parameter space of dark energy models, characterized by periodically modulated potentials such as \refeq{V}, which live outside the standard parametrizations of the equation of state such as \cite{chevallier:00,linder:02}
\be
w_{\rm DE}(a) = w_0 + w_a (a-a_0)\,.
\label{eq:wparam}
\ee
That is, these models show significant, observationally detectable features that are not captured by $(w_0,w_a)$, and require a dedicated search. This is true in particular of models with a non-standard kinetic term, which are not restricted to small oscillations as in the case of quintessence. This search involves scanning the parameter space for periodic modulations with frequency $\nu$ and amplitude $A$ [and in general, a phase $\phi$ which we have set to zero in \refeq{V}] around a slow-rolling background. 
The most promising observables to search for monodromic dark energy are $(i)$ the expansion rate $H$; $(ii)$ the growth rate of structure $f$; and $(iii)$ the amplitude of gravitational lensing, as a function of redshift.

\emph{Hubble rate $H$:} oscillations in the Hubble rate can be probed by taking the derivative of the distance-redshift relation observed with standard candles such as type-Ia Supernovae, for example JLA \cite{JLA:2014} and Supercal \cite{supercal:2015}. Note that the uncertainties involved in the distance ladder are not crucial here, since one is looking for time-dependent features in the Hubble rate. Alternatively, the BAO feature and AP distortions in the two-point function of galaxies and other tracers allow for a measurement of the Hubble rate on a comoving scale of $\sim 110 \Mpch$ in case of the BAO feature, and a wider range of scales for AP distortions. This corresponds to averaging the quantity shown in \reffig{obsbg} over a redshift window of $\Delta z \sim 0.03-0.05$, depending on redshift. Current constraints from BOSS are at the level of approximately 2\% \cite{alam/etal:2016}, which could clearly constrain some of the models presented here in a dedicated analysis.

\emph{Growth rate $f$:} the growth rate is observable through redshift-space distortions. Current constraints on $f \sigma_8$ are at the level of $\sim 10\%$ \cite{Beutler:2012px,Blake:2013nif,delaTorre:2013rpa,sanchez/etal:2016,Beutler:2016arn}. In addition, distance indicators such as Supernovae and disk galaxies through the Tully-Fisher relation yield comparable constraints at $z \lesssim 0.05$ \cite{Turnbull:2011ty,Johnson:2014kaa,huterer/etal:2017}, which could be combined with the results from galaxy clustering at higher redshifts. While the constraints on the growth rate are not as precise as those on the Hubble rate, they can be extended to smaller spatial scales and thus are able to constrain higher frequencies of oscillations. For example, a growth rate measurement on comoving scales of $\sim 20 \Mpch$ corresponds to a redshift interval of $\Delta z \lesssim 0.01$.

\emph{Amplitude of gravitational lensing $\d_{\rm tot}$:} as we have seen in \reffig{obsgrowthtot}, the monodromic k-essence scenario leads to strong oscillatory features in the total energy density perturbation which sources gravitational lensing. Gravitational lensing observables are projected quantities, and the line-of-sight integration strongly suppresses such features in correlations involving lensing alone, such as cosmic shear. However, galaxy-galaxy lensing, the cross-correlation between lensing and source counts in a narrow redshift range, is projected over a much narrower redshift range, and could show observable oscillatory signatures. Galaxy-galaxy lensing has now been measured at high significance in SDSS \cite{mandelbaum/etal:2006}, CFHTLenS \cite{velander/etal:2014}, CFHT imaging of stripe 82 \cite{leauthaud/etal:2017}, in KiDS \cite{KiDS+GAMA}, and DES \cite{DESy1:gg}. The DES year-1 constraints on the amplitude of galaxy-galaxy lensing are better than 10\%.\footnote{This was estimated from the error on the linear bias $b_\times$ in Fig.~14 of \cite{DESy1:gg}.} Hence, this is a further promising probe of the new dark energy phenomenology introduced here. Moreover, combining galaxy-galaxy lensing with probes of $H(z)$ and $f(z)$ will allow for constraints on the speed of sound of dark energy.

Intriguingly, the reconstructed equation of state from a combination of the most recent cosmological data sets derived by \cite{zhao/etal:2017} shows oscillatory features with an amplitude of $\Delta w_\text{DE} \approx 0.6$; similar evidence was previously found from the combination of BOSS BAO measurements in \cite{aubourg/etal:2015}. In the context of the monodromic models introduced here, this could only be explained by the k-essence scenario. In this context, it would be worthwhile to derive the Bayesian evidence for this scenario, which introduces three parameters in addition to the dark energy density. On the other hand, a monodromic k-essence scenario with such a large oscillation amplitude might be in tension with existing galaxy-galaxy lensing data.

Beyond the set of predictions based on linear perturbation theory derived in this paper, interesting phenomenology of monodromic dark energy is expected in the nonlinear regime, in particular for k-essence around the scale of the sound horizon. For example, unusual dynamical effects could appear when the time scale of the oscillations in the dark energy density becomes comparable to the dynamical time of massive halos.

Given this potential source of rich phenomenology in the large-scale structure of the Universe, there is strong motivation to look more deeply into the theoretical constraints outlined in \refsec{thconst}; in particular, the issue of gradient instabilities and whether there are stable monodromic k-essence-type scenarios which do not suffer from a low cutoff or fine-tuning. A promising approach is through the effective field theory of dark energy \cite{creminelli/etal:2009,EFTDE1,EFTDE2}. However, the oscillation period $1/\nu$ in field space adds a new scale, and time translation is no longer weakly broken in monodromic models; rather, it is replaced by a weakly-broken discrete symmetry $\phi \to \phi + 2\pi/\nu$, analogously to the case of axion monodromy inflation (see the discussion in \cite{behbahani/etal:2012}). Further, it would be interesting to find microsopic  scenarios which lead to a potential of the form \refeq{V}, as well as to study natural values for and limits on the period and amplitude of oscillations. We leave these interesting questions as open topics for future work.

\acknowledgments
I am indebted to Eiichiro Komatsu and Masahiro Takada for helpful comments on the draft, and to Eric Linder for discussions and pointing out valuable references.  I further thank Alex Vikman for dicussions on gradient instabilities. 
Finally, I thank the Wissenschaftskolleg zu Berlin, where this paper was completed, for hospitality. 
This work was supported by the Marie Curie Career Integration Grant  (FP7-PEOPLE-2013-CIG) ``FundPhysicsAndLSS,'' and Starting Grant (ERC-2015-STG 678652) ``GrInflaGal'' from the European Research Council.

\bibliographystyle{arxiv_physrev}
\bibliography{REFS}

\def\eprinttmppp@#1arXiv:@{#1}
\providecommand{\arxivlink[1]}{\href{http://arxiv.org/abs/#1}{arXiv:#1}}
\providecommand{\arxivlinknopre[1]}{\href{http://arxiv.org/abs/#1}{#1}}
\providecommand{\eprintmod}[1][XXXX.XXXX]{\IfSubStr{#1}{arXiv}{\arxivlinknopre{#1}}{\arxivlink{#1}}}
\providecommand{\adsurl}[1]{\href{#1}{ADS}}
\begin{thebibliography}{68}
\expandafter\ifx\csname natexlab\endcsname\relax\def\natexlab#1{#1}\fi
\expandafter\ifx\csname bibnamefont\endcsname\relax
  \def\bibnamefont#1{#1}\fi
\expandafter\ifx\csname bibfnamefont\endcsname\relax
  \def\bibfnamefont#1{#1}\fi
\expandafter\ifx\csname citenamefont\endcsname\relax
  \def\citenamefont#1{#1}\fi
\expandafter\ifx\csname url\endcsname\relax
  \def\url#1{\texttt{#1}}\fi
\expandafter\ifx\csname urlprefix\endcsname\relax\def\urlprefix{URL }\fi

\bibitem{Riess:1998cb}
Supernova Search Team, A.~G. Riess {\em et~al.},
\newblock Astron. J. {\bf 116}, 1009 (1998), [\eprintmod[astro-ph/9805201]].

\bibitem{Perlmutter:1998np}
Supernova Cosmology Project, S.~Perlmutter {\em et~al.},
\newblock Astrophys. J. {\bf 517}, 565 (1999), [\eprintmod[astro-ph/9812133]].

\bibitem{Frieman:2008sn}
J.~Frieman, M.~Turner and D.~Huterer,
\newblock Ann. Rev. Astron. Astrophys. {\bf 46}, 385 (2008),
  [\eprintmod[0803.0982]].

\bibitem{JLA:2014}
M.~{Betoule} {\em et~al.},
\newblock A\&A {\bf 568}, A22 (2014), [\eprintmod[1401.4064]].

\bibitem{planck:2015}
{Planck Collaboration} {\em et~al.},
\newblock A\&A {\bf 594}, A13 (2016), [\eprintmod[1502.01589]].

\bibitem{alam/etal:2016}
S.~{Alam} {\em et~al.},
\newblock \mnras {\bf 470}, 2617 (2017), [\eprintmod[1607.03155]].

\bibitem{DESy1}
{DES Collaboration} {\em et~al.},
\newblock ArXiv e-prints  (2017), [\eprintmod[1708.01530]].

\bibitem{huterer/shafer:2017}
D.~{Huterer} and D.~L. {Shafer},
\newblock ArXiv e-prints  (2017), [\eprintmod[1709.01091]].

\bibitem{Weinberg:1988cp}
S.~Weinberg,
\newblock Rev. Mod. Phys. {\bf 61}, 1 (1989).

\bibitem{Copeland:2006wr}
E.~J. Copeland, M.~Sami and S.~Tsujikawa,
\newblock Int. J. Mod. Phys. {\bf D15}, 1753 (2006),
  [\eprintmod[hep-th/0603057]].

\bibitem{joyce/lombriser/schmidt:2016}
A.~{Joyce}, L.~{Lombriser} and F.~{Schmidt},
\newblock Annual Review of Nuclear and Particle Science {\bf 66}, 95 (2016),
  [\eprintmod[1601.06133]].

\bibitem{Wetterich:1987fm}
C.~Wetterich,
\newblock Nucl. Phys. {\bf B302}, 668 (1988).

\bibitem{Peebles:1987ek}
P.~J.~E. Peebles and B.~Ratra,
\newblock Astrophys. J. {\bf 325}, L17 (1988).

\bibitem{ratra/peebles}
B.~Ratra and P.~J.~E. Peebles,
\newblock Phys. Rev. D {\bf 37}, 3406 (1988).

\bibitem{silverstein/westphal:2008}
E.~{Silverstein} and A.~{Westphal},
\newblock \prd {\bf 78}, 106003 (2008), [\eprintmod[0803.3085]].

\bibitem{mcallister/etal:2010}
L.~{McAllister}, E.~{Silverstein} and A.~{Westphal},
\newblock \prd {\bf 82}, 046003 (2010), [\eprintmod[0808.0706]].

\bibitem{chen/easther/lim}
X.~{Chen}, R.~{Easther} and E.~A. {Lim},
\newblock \jcap {\bf 4}, 010 (2008), [\eprintmod[0801.3295]].

\bibitem{flauger/etal:2010}
R.~{Flauger}, L.~{McAllister}, E.~{Pajer}, A.~{Westphal} and G.~{Xu},
\newblock \jcap {\bf 6}, 009 (2010), [\eprintmod[0907.2916]].

\bibitem{flauger/pajer}
R.~{Flauger} and E.~{Pajer},
\newblock \jcap {\bf 1}, 017 (2011), [\eprintmod[1002.0833]].

\bibitem{dodelson/kaplinghat/stewart:2000}
S.~{Dodelson}, M.~{Kaplinghat} and E.~{Stewart},
\newblock Physical Review Letters {\bf 85}, 5276 (2000),
  [\eprintmod[astro-ph/0002360]].

\bibitem{frieman/etal:1995}
J.~A. {Frieman}, C.~T. {Hill}, A.~{Stebbins} and I.~{Waga},
\newblock Physical Review Letters {\bf 75}, 2077 (1995),
  [\eprintmod[astro-ph/9505060]].

\bibitem{kamionkowski/pradler/walker}
M.~{Kamionkowski}, J.~{Pradler} and D.~G.~E. {Walker},
\newblock Physical Review Letters {\bf 113}, 251302 (2014),
  [\eprintmod[1409.0549]].

\bibitem{damico/etal:2016}
G.~{D'Amico}, T.~{Hamill} and N.~{Kaloper},
\newblock \prd {\bf 94}, 103526 (2016), [\eprintmod[1605.00996]].

\bibitem{spintessence1}
J.-A. Gu and W.-Y.~P. Hwang,
\newblock Phys. Lett. {\bf B517}, 1 (2001), [\eprintmod[astro-ph/0105099]].

\bibitem{spintessence2}
L.~A. Boyle, R.~R. Caldwell and M.~Kamionkowski,
\newblock Phys. Lett. {\bf B545}, 17 (2002), [\eprintmod[astro-ph/0105318]].

\bibitem{spintessence3}
V.~Sahni and L.-M. Wang,
\newblock Phys. Rev. {\bf D62}, 103517 (2000), [\eprintmod[astro-ph/9910097]].

\bibitem{spintessence4}
S.~Dutta and R.~J. Scherrer,
\newblock Phys. Rev. D {\bf 78}, 083512 (2008).

\bibitem{johnson/kamionkowski}
M.~C. Johnson and M.~Kamionkowski,
\newblock Phys. Rev. {\bf D78}, 063010 (2008), [\eprintmod[0805.1748]].

\bibitem{barenboim1}
G.~Barenboim and J.~Lykken,
\newblock Physics Letters B {\bf 633}, 453  (2006).

\bibitem{barenboim2}
G.~Barenboim, O.~M. Requejo and C.~Quigg,
\newblock Journal of Cosmology and Astroparticle Physics {\bf 2006}, 008
  (2006).

\bibitem{linder:2006}
E.~V. {Linder},
\newblock Astroparticle Physics {\bf 25}, 167 (2006),
  [\eprintmod[astro-ph/0511415]].

\bibitem{kurek2}
A.~Kurek, O.~Hrycyna and M.~Szydłowski,
\newblock Physics Letters B {\bf 690}, 337  (2010).

\bibitem{xia/etal:2005}
J.-Q. Xia, B.~Feng and X.~Zhang,
\newblock Modern Physics Letters A {\bf 20}, 2409 (2005).

\bibitem{kurek1}
A.~Kurek, O.~Hrycyna and M.~Szydłowski,
\newblock Physics Letters B {\bf 659}, 14  (2008).

\bibitem{pace/etal:2011}
F.~Pace, C.~Fedeli, L.~Moscardini and M.~Bartelmann,
\newblock Monthly Notices of the Royal Astronomical Society {\bf 422}, 1186
  (2012).

\bibitem{chiba/okabe/yamaguchi}
T.~{Chiba}, T.~{Okabe} and M.~{Yamaguchi},
\newblock \prd {\bf 62}, 023511 (2000), [\eprintmod[astro-ph/9912463]].

\bibitem{ArmendarizPicon:2000ah}
C.~Armendariz-Picon, V.~F. Mukhanov and P.~J. Steinhardt,
\newblock Phys. Rev. {\bf D63}, 103510 (2001), [\eprintmod[astro-ph/0006373]].

\bibitem{silverstein/tong}
E.~{Silverstein} and D.~{Tong},
\newblock \prd {\bf 70}, 103505 (2004), [\eprintmod[hep-th/0310221]].

\bibitem{alishahiha/etal:2004}
M.~{Alishahiha}, E.~{Silverstein} and D.~{Tong},
\newblock \prd {\bf 70}, 123505 (2004), [\eprintmod[hep-th/0404084]].

\bibitem{deputter/linder:2008}
R.~{de Putter} and E.~V. {Linder},
\newblock \jcap {\bf 10}, 042 (2008), [\eprintmod[0808.0189]].

\bibitem{creminelli/etal:2009}
P.~{Creminelli}, G.~{D'Amico}, J.~{Nore{\~n}a} and F.~{Vernizzi},
\newblock \jcap {\bf 2}, 018 (2009), [\eprintmod[0811.0827]].

\bibitem{vikman:2004}
A.~Vikman,
\newblock Phys. Rev. {\bf D71}, 023515 (2005), [\eprintmod[astro-ph/0407107]].

\bibitem{kaiser:1987}
N.~Kaiser,
\newblock Monthly Notices of the Royal Astronomical Society (ISSN 0035-8711)
  {\bf 227}, 1 (1987).

\bibitem{bean/dore}
R.~{Bean} and O.~{Dor{\'e}},
\newblock \prd {\bf 69}, 083503 (2004), [\eprintmod[astro-ph/0307100]].

\bibitem{sefusatti/vernizzi}
E.~Sefusatti and F.~Vernizzi,
\newblock JCAP {\bf 1103}, 047 (2011), [\eprintmod[1101.1026]].

\bibitem{takada:2006}
M.~{Takada},
\newblock \prd {\bf 74}, 043505 (2006), [\eprintmod[astro-ph/0606533]].

\bibitem{chevallier:00}
M.~Chevallier and D.~Polarski,
\newblock Int. J. Mod. Phys. {\bf D10}, 213 (2001),
  [\eprintmod[gr-qc/0009008]].

\bibitem{linder:02}
E.~V. Linder,
\newblock Phys. Rev. Lett. {\bf 90}, 091301 (2003),
  [\eprintmod[astro-ph/0208512]].

\bibitem{supercal:2015}
D.~{Scolnic} {\em et~al.},
\newblock \apj {\bf 815}, 117 (2015), [\eprintmod[1508.05361]].

\bibitem{Beutler:2012px}
F.~Beutler {\em et~al.},
\newblock Mon. Not. Roy. Astron. Soc. {\bf 423}, 3430 (2012),
  [\eprintmod[1204.4725]].

\bibitem{Blake:2013nif}
C.~Blake {\em et~al.},
\newblock Mon. Not. Roy. Astron. Soc. {\bf 436}, 3089 (2013),
  [\eprintmod[1309.5556]].

\bibitem{delaTorre:2013rpa}
S.~de~la Torre {\em et~al.},
\newblock Astron. Astrophys. {\bf 557}, A54 (2013), [\eprintmod[1303.2622]].

\bibitem{sanchez/etal:2016}
A.~G. {S{\'a}nchez} {\em et~al.},
\newblock \mnras {\bf 464}, 1640 (2017), [\eprintmod[1607.03147]].

\bibitem{Beutler:2016arn}
BOSS, F.~Beutler {\em et~al.},
\newblock Submitted to: Mon. Not. Roy. Astron. Soc.  (2016),
  [\eprintmod[1607.03150]].

\bibitem{Turnbull:2011ty}
S.~J. Turnbull {\em et~al.},
\newblock Mon. Not. Roy. Astron. Soc. {\bf 420}, 447 (2012),
  [\eprintmod[1111.0631]].

\bibitem{Johnson:2014kaa}
A.~Johnson {\em et~al.},
\newblock Mon. Not. Roy. Astron. Soc. {\bf 444}, 3926 (2014),
  [\eprintmod[1404.3799]].

\bibitem{huterer/etal:2017}
D.~{Huterer}, D.~L. {Shafer}, D.~M. {Scolnic} and F.~{Schmidt},
\newblock \jcap {\bf 5}, 015 (2017), [\eprintmod[1611.09862]].

\bibitem{mandelbaum/etal:2006}
R.~{Mandelbaum}, U.~{Seljak}, G.~{Kauffmann}, C.~M. {Hirata} and
  J.~{Brinkmann},
\newblock \mnras {\bf 368}, 715 (2006), [\eprintmod[astro-ph/0511164]].

\bibitem{velander/etal:2014}
M.~{Velander} {\em et~al.},
\newblock \mnras {\bf 437}, 2111 (2014), [\eprintmod[1304.4265]].

\bibitem{leauthaud/etal:2017}
A.~{Leauthaud} {\em et~al.},
\newblock \mnras {\bf 467}, 3024 (2017), [\eprintmod[1611.08606]].

\bibitem{KiDS+GAMA}
E.~{van Uitert} {\em et~al.},
\newblock ArXiv e-prints  (2017), [\eprintmod[1706.05004]].

\bibitem{DESy1:gg}
J.~{Prat} {\em et~al.},
\newblock ArXiv e-prints  (2017), [\eprintmod[1708.01537]].

\bibitem{zhao/etal:2017}
G.-B. {Zhao} {\em et~al.},
\newblock ArXiv e-prints  (2017), [\eprintmod[1701.08165]].

\bibitem{aubourg/etal:2015}
{\'E}.~{Aubourg} {\em et~al.},
\newblock \prd {\bf 92}, 123516 (2015), [\eprintmod[1411.1074]].

\bibitem{EFTDE1}
G.~{Gubitosi}, F.~{Piazza} and F.~{Vernizzi},
\newblock \jcap {\bf 2}, 032 (2013), [\eprintmod[1210.0201]].

\bibitem{EFTDE2}
J.~{Bloomfield}, {\'E}.~{\'E}. {Flanagan}, M.~{Park} and S.~{Watson},
\newblock \jcap {\bf 8}, 010 (2013), [\eprintmod[1211.7054]].

\bibitem{behbahani/etal:2012}
S.~R. {Behbahani}, A.~{Dymarsky}, M.~{Mirbabayi} and L.~{Senatore},
\newblock \jcap {\bf 12}, 036 (2012), [\eprintmod[1111.3373]].

\end{thebibliography}

\end{document}